\documentclass[11pt, a4paper]{article} 
\usepackage[utf8]{inputenc}
\usepackage[left=2cm, right=2cm, top=2.1cm, bottom=2.1cm]{geometry} 

\usepackage{enumitem} 
\usepackage{textgreek} 
\usepackage{physics} 
\usepackage{amsmath}
\usepackage{amsthm} 
\usepackage{amssymb}
\usepackage{mathtools} 
\usepackage[scr=boondoxo]{mathalfa}
\usepackage{pst-node}%
\usepackage{mathrsfs}
\DeclareMathAlphabet{\mathpzc}{OT1}{pzc}{m}{it}

\newcommand{\R}{\mathbb{R}} 

\newcommand{\B}{\mathfrak{B}}

\usepackage{csquotes}
\MakeOuterQuote{"}
\usepackage[%
style=phys,backend=bibtex,
articletitle=false,biblabel=brackets,%
chaptertitle=false,pageranges=false
]
{biblatex}
\usepackage{graphicx} 
\graphicspath{{Figures/}} 
\usepackage{subcaption} 

\usepackage{caption}
\usepackage{booktabs} 
\usepackage{siunitx} 
\usepackage{hyperref} 
\hypersetup{
    colorlinks=true,
    linkcolor=violet,
    filecolor=violet,      
    urlcolor=violet,
    citecolor=violet
}

\usepackage{etoolbox}
\makeatletter
\patchcmd{\epigraph}{\@epitext{#1}}{\itshape\@epitext{#1}}{}{}

\title{}
\date{\vspace{-11ex}}

\usepackage{adjustbox}
\newcolumntype{?}{!{\vrule width 1.5pt}}
\usepackage{abstract}
\usepackage{anyfontsize}

\AtBeginBibliography{\vspace*{-0.2cm}}
\DeclareMathSizes{10}{10}{10}{10}

\begin{document}
\vspace{1cm}

\begin{center}
\section*{ Bohmian Mechanics as a Practical Tool \vspace{0.4cm}\vspace{0.1cm}\\ \small Xabier Oianguren-Asua{\normalsize \footnote{\em Departament de Física, Universitat Autònoma de Barcelona, 08193 Bellaterra, Barcelona, Spain}}, Carlos F. Destefani{\normalsize\footnote{\em Departament d’Enginyeria Electrònica, Universitat Autònoma de Barcelona, 08193 Bellaterra, Barcelona, Spain}}, Matteo Villani{\normalsize$^2$}, David K. Ferry{\normalsize\footnote{\em School of Electrical, Computer, and Energy Engineering, Arizona State University, Tempe,
AZ 85287 USA}}, Xavier Oriols{\normalsize$^2$}}

\vspace{0.6cm}
{\bf \small  Chapter of\vspace{0.3cm}\\ A. Bassi et al. (eds.), {\em Physics and the Nature of Reality: Essays in Memory of Detlef Dürr}, Fundamental Theories
of Physics 215. Springer (2024).\footnote{\href{https://doi.org/10.1007/978-3-031-45434-9_9}{DOI : $10.1007/978-3-031-45434-9\_9$}}}\vspace{-0.32cm}
\end{center}
\vspace{1cm}

\begin{abstract}
\vspace{-0.2cm}
\hspace{5mm}\normalsize In this chapter, we will take a trip around several hot-spots where Bohmian
mechanics and its capacity to describe the microscopic reality, even in the absence
of measurements, can be harnessed as computational tools, in order to help in the
prediction of phenomenologically accessible information (also useful for the followers of the Copenhagen theory). As a first example, we will see how a Stochastic
Schrödinger Equation, when used to compute the reduced density matrix of a non Markovian open quantum system, necessarily seems to employ the Bohmian concept
of a conditional wavefunction. We will see that by dressing these conditional wavefunctions with an interpretation, the Bohmian theory can prove to be a useful tool
to build general quantum simulation frameworks, such as a high-frequency electron
transport model. As a second example, we will explain how a Copenhagen “observable operator” can be related to numerical properties of the Bohmian trajectories,
which within Bohmian mechanics, are well-defined even for an “unmeasured” system. Most importantly in practice, even if these numbers are given no ontological
meaning, not only we will be able to simulate (thus, predict and talk about) them,
but we will see that they can be operationally determined in a weak value experiment. Therefore, they will be practical numbers to characterize a quantum system
irrespective of the followed quantum theory.
\end{abstract}

\vspace{1cm}

\pagenumbering{gobble}

\newpage

\pagenumbering{arabic}
\setcounter{page}{1}

\subsection*{1. Introduction}\vspace{-0.2cm}
Questioning whether “there are” electrons inside our mobile phones sounds like an
absurd reflection, and yet the standard (also called Copenhagen or orthodox) quantum theory is not able to affirm it \cite{where, consp}.  Under this theory, a quantum object has a
well-defined property (like the position) only when its wavefunction is an eigenstate
of the associated operator. We know that this happens when the property is “strongly
measured”. But in general, the wavefunction is in a superposition of eigenstates for
the operator of that observable, meaning nothing can be said about it: the property becomes “unspeakable” until measured. Consequently, the Copenhagen theory
affirms that it is meaningless to talk about, say, the positions of the electrons inside
the active regions of nanoscale devices, because while in operation, their position
is never (strongly) measured. Thus, there is no chance for an affirmative answer to
our initial question. And yet, consciously or not, no engineer or applied physicist
can seriously accept there is no electron in an operating nano-device like a transistor \cite{where, consp}. Remarkably, alternatives to the Copenhagen interpretation of quantum mechanics exist, by which electrons have a defined position irrespective of their measurement
and the state of superposition of their wavefunction, e.g., the well-known Bohmian
interpretation \cite{Bohm,Holland, Durr,JordiXavier}. \vspace{-0.1cm} 

What might be more relevant from a practical point of view, however, is that even
if one turns a blind eye to these “picky unspeakabilities” of the Copenhagen theory,
their implications also limit the employable modelling tool-set, making some scenarios look (unnecessarily) pathological. For example, the explained undefined position
of electrons comes into conflict with a well-defined dwell time for the electrons in the
active region of a nano-scale transistor, which is an essential parameter to predict the
performance of next generation computers. Similar practical issues can be found in
the search of measurement operators (like the multi-electron displacement current \cite{equiv, Pel}) in scenarios where their mathematical shape is non-trivial (e.g., in nano-scale
devices operating at THz frequencies \cite{Thz}), or when looking for pure state “unravellings” in non-Markovian open quantum systems. As we will see in this chapter, such
problems happen to be unambiguously solvable under the Bohmian quantum theory.  \vspace{-0.2cm}

\subsubsection*{1.1. A Suggestive Review}
\vspace{-0.2cm}
We can arrive at these conclusions through the inherently Bohmian concepts of a
conditional wavefunction (CWF) and an effective wavefunction (EWF), introduced by Dürr et al. \cite{Absolute}, together with the understanding of the measurement dilemma they
illuminate \cite{operatorsObservables}. But before going into the details, let us note that only {\em non-relativistic quantum phenomena} will be discussed in this chapter. The spirit is to show that,
for such phenomena and their formulation, the Bohmian theory provides a
most convenient narrative.

Given an isolated quantum system of $N$ degrees of freedom described by the real coordinate vector $\vec{q}=(q_1,..., q_N)\in\R^N$, its evolution in time $t$ within Bohmian mechanics is given by two maps: a complex wavefunction $\Psi(\vec{q},t)$, which in polar form $\Psi(\vec{q},t)=\rho^{1/2}(\vec{q},t)e^{i\mathcal{S}(\vec{q},t)/\hbar}$ encodes the real fields $\mathcal{S}$ and $\rho$, and a trajectory $\vec{q}^{\:\xi}(t)\equiv \vec{q}\:(\vec{\xi},t)$ for the degrees of freedom of the system, the initial condition of which, $\vec{q}^{\:\xi}(t=0)=\vec{\xi}\in \R^N$, labels the actual trajectory among the possible ones. This trajectory is piloted by the
wavefunction, which provides the velocity field $v_k$ for the $k$-th degree of freedom as $v_k(\vec{q},t):=\frac{1}{m_k} \pdv{\mathcal{S}(\vec{q},t)}{x_k}$ \cite{Bohm,Holland,Durr,JordiXavier}. Meanwhile, the wavefunction itself is guided by the Schrödinger Equation\vspace{-0.1cm} 
\begin{equation}\label{SE}
i\hbar\pdv{\Psi(\vec{q},t)}{t}=\Big[ \sum_{k=1}^N \frac{-\hbar^2}{2m_k}\pdv[2]{}{q_k}+U(\vec{q})\Big]\Psi(\vec{q},t),\vspace{-0.1cm}
\end{equation}
where $m_k$ is the mass associated with the $k$-th degree of freedom and $U$ denotes the potential energy field describing the interaction between the degrees of freedom. 

The most general isolated system we can consider is the entire Universe, where $\vec{q}$ would reflect its possible {\em configurations}. This system can be partitioned into a subsystem of interest S, of $n<N$ degrees of freedom $\vec{x}=(x_1,...,x_n)$, and its environment, of degrees of freedom $\vec{y}=(y_{n+1},...,y_N)$, such that $\vec{q}\equiv (\vec{x},\vec{y})$. Bohmian
mechanics allows us to associate to the system and the environment their own wavefunctions, labelled by the initial joint configuration $\vec{\xi}$, as $\psi^\xi(\vec{x},t):=\Psi(\vec{x},\vec{y}^{\:\xi}(t),t)$ and $\varphi^\xi(\vec{y},t):=\Psi(\vec{x}^{\:\xi}(t),\vec{y},t)$. These are particular cases of the so called {\em conditional wavefunctions}. In general, a CWF is a “slice” of a wavefunction, obtained
by evaluating some of its degrees of freedom along a (Bohmian) trajectory, while
leaving the rest of them un-evaluated \cite{Absolute, JordiXavier}. Now, a priori, the actual trajectories $\vec{x}^{\:\xi}(t)$ and $\vec{y}^{\:\xi}(t)$ are unknown, but by the Quantum Equilibrium principle \cite{Absolute}, if the trajectory of the whole Universe had a "typical" initial condition $\vec{\xi}$, the probability density of the position $\vec{x}$ at time $t$ (resp. $\vec{y}$), will be given by the CWF as $|\psi^\xi(\vec{x},t)|^2$ (resp. $|\varphi^\xi(\vec{y},t)|^2$).

As proved in \cite{GJ}, the full Schrödinger Equation \eqref{SE} can be rewritten exactly into
a coupled pair of dynamical equations ruling the motion of the two CWFs. Assuming we can write $U(\vec{x},\vec{y}\,)=U_x(\vec{x}\,)+U_{xy}(\vec{x},\vec{y}\,)$, for the system S we have\footnote{ For the environment the equation will be the same but changing the CWF and the index ranges in \eqref{SE.GJ} and \eqref{G.Bohm}.}\vspace{-0.1cm}
\begin{equation}\label{SE.GJ}
i \hbar \pdv{\psi^\xi (\vec{x},t)}{t} = \qty[\sum_{k=1}^n\frac{-\hbar^2}{2m_k} \pdv[2]{}{x_k} + U_x(\vec{x}\,)+ U_{xy}(\vec{x}, \vec{y}^{\: \xi}(t)) + \mathfrak{W}(\vec{x}, \vec{y}^{\: \xi}(t),t)] \psi^\xi (\vec{x},t),
\end{equation}
where $\mathfrak{W}$ is the so-called {\em quantum correlation potential}
   \begin{equation}\label{G.Bohm}
      \mathfrak{W}(\vec{x},\vec{y}^{\:\xi}(t),t):=\sum_{j=n+1}^N\Bigg[-\frac{\hbar^2}{2m_j\rho^{1/2}}\Bigg(\frac{\partial^2\rho^{1/2}(\vec{x},\vec{y},t)}{\partial y_j^2}\Bigg) -\frac{1}{2}m_jv_j^2(\vec{x},\vec{y},t)-i\frac{\hbar}{2}\frac{\partial v_j(\vec{x},\vec{y},t)}{\partial y_j}\Bigg]\Big\rvert_{\vec{y}=\vec{y}^{\:\xi}(t)},
   \end{equation}
where we recognize as its real part $\R e\{\mathfrak{W}\}$ the difference between the Bohmian quantum potential \cite{JordiXavier, Durr} and the kinetic energy of the environment degrees of freedom $y_j$; and as the imaginary part $\mathbb{I}m\{\mathfrak{W}\}$, the spatial variation in the environment axes $y_j$ of their associated Bohmian velocity. The evaluation of both parts involves, at each $\vec{x}$, a derivative of the phase $\mathcal{S}$ or of the magnitude $\rho$ of the full wavefunction $\Psi$ along the environment coordinates $\vec{y}$, centered at the trajectory position $\vec{y}^{\:\xi}(t)$. This means $\mathfrak{W}$ requires information about the wavefunction over nearby possible trajectories $\vec{y}^{\:\xi'}(t)=\vec{y}^{\:\xi}(t)+\Delta \vec y$, with $|\Delta\vec y\, |$ small. That is, the evolution of the CWF $\psi^\xi (\vec{x},t)$ depends on other adjacent CWFs or slices of the full wavefunction (with different $\vec{\xi}$). This feature is known as "quantum wholeness" \cite{JordiXavier}.

Now, we might ask when the subsystem CWF $\psi^\xi(\vec{x},t)$ behaves as if it was the wavefunction of a quantum system that is independent of the environment, i.e., ruled by a Schrödinger Equation like Eq. \eqref{SE}. We see in Eq. \eqref{SE.GJ} that this happens for instance when $\mathfrak{W}=f(t)$ (adding only a global phase) and $U_{xy}(\vec{x},\vec{y}^{\,\xi}(t))\simeq V(\vec{x},t)$ with a same shape irrespective of the trajectory $\vec{\xi}$.\footnote{ If only $\mathbb{I}m\{\mathfrak{W}\}$ vanished, the CWF would already seem to be ruled by a unitary Schrödinger Equation of a closed system, with a real potential energy field defined as $V(\vec{x},t):=U(\vec{x},\vec{y}^{\:\xi}(t))+\R e\{\mathfrak{W}(\vec{x},\vec{y}^{\:\xi}(t),t)\}$. However, computationally, in order to evaluate $\R e\{\mathfrak{W}\}$ and the trajectory $\vec{y}^\xi(t)$, a quantum description of the environment would still be required, making the CWF of S not independent of the environment's evolution and thus, not an EWF. } Whenever this is the case, one says that the CWF of the system is its {\em effective wavefunction}. The question is then: when do these two conditions happen? One of the most important cases is just after a "strong measurement" of the subsystem. 

This is well-known in the Bohmian literature about measurement \cite{Durr, Absolute, operatorsObservables}, but let us review it qualitatively, because it will be key to understand Markovianity. Let there be an initially closed quantum system S with EWF $|\psi(0)\rangle_S=\int\psi^\xi(\vec{x},t=0)|\vec{x}\rangle d^nx$. As part of the environment of S, let us consider the degree of freedom of the pointer of a macroscopic measuring apparatus M, $z\equiv y_{n+1}$, following the standard von Neumann protocol \cite{vonNeumann}. Initially this pointer will be around its repose position, independently of the rest of the environment, meaning it should have a  spatially localized EWF $\ket{\varphi(0)}_M$. In order to transfer information from S to M, they are made to interact until $t=T$, through the von Neumann coupling Hamiltonian $\hat{H}_{MS}:=\bar{\mu}(t)\,\hat{p}_M\otimes \hat{B}_S$, where $\hat{p}_M$ is the pointer's momentum operator, $\hat{B}_S=\sum_kb_k\ket{b_k}\bra{b_k}$ with $b_k\in\R$ is the "diagonalized" self-adjoint operator related to the system's property $B$ to be "measured" and $\mu:=\int_0^T\bar{\mu}(t)dt$ is the interaction strength. This Hamiltonian entangles the position of the pointer $z$ with the eigenstates $\ket{b_k}_S$ of $\hat{B}_S$ such that, the system-pointer wavefunction will separate the eigenstates $\ket{b_k}_S$ along the configuration space axis $z$, by enveloping each with a differently displaced version of the localized $\ket{\varphi(0)}_M$, scaled by $P_k:=|\bra{b_k}_S\ket{\psi}_S|^2$. For a reproducible measurement \cite{operatorsObservables} and the pointer to show us macroscopically distinguishable results $b_k$, the interaction $\mu$, proportional to the separation of the envelopes, must be strong enough to leave them macroscopically disjoint in $z$. Then, the pointer will show a position $z^\xi(T)$ around one of these envelopes, which will "slice" a CWF for S equal to the eigenstate $\ket{b_j}_S$ modulated by that envelope. This will happen with probability $P_j$ (area of the envelope) given by the Quantum Equilibrium principle \cite{Absolute}. The CWFs linked to the rest of possible envelopes are called "empty waves". At this point, the Copenhagen theory postulates a so-called "wavefunction collapse", that transforms the entangled wavefunction into a product of a single eigenstate $\ket{b_j}_S$ and its corresponding envelope \cite{vonNeumann}. In Bohmian mechanics, there is no need to postulate any physical "collapse", instead it is seen just as an apparent process caused by the effective vanishing of the correlation potential $\mathfrak{W}$ for S. This vanishing is due to the fact that the different CWF "sets" enveloped along $z$ have a macroscopically disjoint support and because for all relevant $t>T$, the system-apparatus coupling potential vanishes ($\bar{\mu}(t)=0$). In consequence, the CWF for S selected by the Bohmian position of the pointer will evolve for $t>T$ as if it were again an independent closed quantum system wavefunction: it will be an EWF. Since the EWF is enough for the complete future description of S, we can omit the description of the measurement process that took an (ideally small) time $T$ and consider an "effective collapse" $\ket{\psi}_S\rightarrow\ket{b_j}_S$.

Notice that either the assumption that, for time $t>T$, M does not interact anymore with S, or that its entanglement with S is lost by some sort of thermalisation (by which the empty waves get macroscopically dispersed in configuration space \cite{Absolute}), mean that the information of S "leaked" to the environment M, the "empty waves", do not interact back with the EWF of S. Therefore, these assumptions imply that the environment effectively "forgets" the entanglement achieved with S. This is an environment behaviour we could call memory-less or {\em Markovian}. 

Using this effective collapse idea, we can extract more general information about the subsystem. If part of the environment, let us call it the "ancilla" A, gets entangled with S and this ancilla then suffers an effective collapse as in the measurement we just described, S will also suffer an effective "collapse", but now into non-necessarily orthogonal, nor linearly independent states. Let $\ket{\theta_0}_A$ and $\ket{\psi}_S$ be the EWFs of A and S before their interaction, then a general unitary evolution coupling them will yield $\hat{U}_{AS}\ket{\theta_0}_A\otimes\ket{\psi}_S=\sum_m \ket{\theta_m}_A\otimes \hat{M}_m\ket{\psi}_S$, with $\{\ket{\theta_m}_A\}_m$ an orthonormal basis of A's Hilbert space and $\{\hat{M}_m\}_m$ a family of bounded linear operators on S, such that $\sum_m\hat{M}_m^\dagger\hat{M}_m$ equals the identity. By measuring the observable of A with eigenstates $\{\ket{\theta_m}_A\}_m$, the composite will collapse into the (unnormalized) EWFs $\ket{\theta_m}\otimes \hat{M}_m\ket{\psi}_S$, with probabilities $P_m:= \langle \psi|_S\hat{M}_m^\dagger \hat{M}_m|\psi\rangle_S$, where the corresponding CWF of S would be $\ket{\phi_m}_S:=\hat{M}_m\ket{\psi}_S$. If A and S stop interacting, $\ket{\phi_m}_S$ will become possible EWFs of S, called {\em "conditional states"} (which are a particular types of CWFs).\footnote{If the measurement was for the position operator of A, $\ket{\phi_m}_S$ would be the CWFs of the system for the state $\hat{U}_{AS}\ket{\theta_0}_A\otimes\ket{\psi}_S$ as it was {\em before the strong measurement of A}, otherwise, they would only be CWFs of the collapsed $\ket{\theta_m}_A\otimes \hat{M}_m\ket{\psi}_S$.} This process is called a {\em generalized measurement} of S \cite{Generalized, operatorsObservables}.

Within the described scenario, consider the interpretation of density matrices in Bohmian mechanics, as useful tools for statistical predictions about stochastic ensembles of wavefunctions (even if they provide an incomplete microscopic description) \cite{density, operatorsObservables}. It turns out that, the partial trace of A in the state $\hat{U}_{AS}\ket{\theta_0}_A\otimes\ket{\psi}_S$ yields the {\em unconditional} post-measurement density matrix $\hat{\rho}_S=\sum_m \ket{\phi_m}_S\bra{\phi_m}_S$, which is the density matrix that keeps track of all possible measurement results. Now, for any state $|\Psi\rangle_{AS}$ of the composite there exists a unitary operator $U_{AS}$ such that $|\Psi\rangle_{AS}= U_{AS} |\theta_0\rangle_A \otimes |\psi\rangle_S$. This suggests, and turns out to be the case, that the partial trace of an ancilla partition A of a composite AS state, called, the {\em reduced density matrix of S,} can always be interpreted as an unconditional post-(generalized) measurement density matrix. That is, as how an ensemble of identical subsystems S, each coupled to an identical ancilla A, would be left if an strong measurement was performed on each ancilla A \cite{Generalized}. Now, by the uniqueness of the partial trace, this "fictitious measurement" of A could be considered to be for an arbitrary observable. For example, since we could choose the position operator of A, the reduced density matrix of S can always be computed by the ensemble average over possible CWF-s. The key remark after this result is that if the traced partition A is not really measured at $t$ and the entanglement between A and S is not "thermalised" and their interaction does not cease indeterminately, then the reduced density matrix of S will just be a "fiction" if interpreted as describing independent possible quantum states. Each conditional state of S, each $\ket{\phi_m}_S$, will still interact  through the environment's degrees of freedom with each other, namely, they will not be (unnormalized) EWFs. Therefore, even if the {\em reduced} density matrix is enough to predict measurement statistics on S, in typical scenarios it will not convey enough information to predict its time evolution. The traced out environment's entanglement with the subsystem will need to be tracked for that. This could then be called an environment with memory, or {\em non-Markovian environment}. For example, the microscopic information about the spatial distribution of the CWFs along the environment's axes would be required to predict the time evolution of the reduced density matrix, which is encapsulated in the correlation potential \eqref{G.Bohm} (or in the so-called "memory time superoperator" \cite{WisemanSSE} of the standard quantum theory). To know this without explicit simulation of the environment is precisely the challenge of open quantum systems.\vspace{-0.15cm}

\subsection*{2. How Markovian and non-Markovian Stochastic Schrödinger Equations tacitly employ Conditional Wavefunctions}\vspace{-0.05cm}
There are scenarios where the post-measurement density matrix "fiction" developed above does provide a reasonable description of a reduced density matrix. Consider, for instance, a scenario where every $\Delta t$ time units, a different portion of the environment (a different ancilla) got coupled with S and was then (the ancilla) strongly measured. If these ancillas never again interacted with S (or their entanglement was somehow "thermalised" before their next interaction with S), the result would be equivalent to a generalized measurement of S every $\Delta t$. Following our previous comments, we could call such a system S, a Markovian open quantum system \cite{QuantumTrajs}, satisfying, among others, the "Past-Future Independence" characterization of Ref. \cite{MarkovianityDefs}. Remarkably, as shown by Ref. \cite{continousMeas}, if $\Delta t\rightarrow 0$ such a continuous measurement of ancillas that sequentially get coupled to the subsystem S, can be used to derive the dynamical equation for the reduced density matrix of S (called, the {\em master equation}) for {\em many} relevant Markovian settings. What is more, it is proven that for {\em any} Markovian master equation a (perhaps fictitious) environment and a set of observables for it exist, such that the equation is interpretable as due to their simultaneous and continuous measurements \cite{continousMeas, MarkovianityDefs}.

As a consequence, for a Markovian environment, instead of directly solving the master equation for S, the next procedure is possible. First, find (fictitious or not) environmental ancillas and observables $W$, such that if the ancillas got entangled with S one after the other, and their properties $W$ were sequentially measured, they would cause the same (unconditional) evolution of the reduced density matrix of S, as the one described by the master equation (which in principle is possible for all Markovian master equations). Then, if a pure state-vector of S was evolved, by choosing for each projective measurement of the bath ancillas, one of the possible post-measurement conditional states, this would generate pure states $\ket{\psi_{w(t)}(t)}_S$, associated with a certain chain of measurement results (an unravelling) for the bath ancillas: $w(t)$.\footnote{At each time a different generalized measurement is performed on S, meaning the stochastic trajectory $w(t)$ reflects the Bohmian positions of different measurement pointers at each $\Delta t$. Its non-differentiability is thus unproblematic.} This pure state $\ket{\psi_{w(t)}(t)}_S$ is called the "quantum trajectory" linked to the "noise realization" $w(t)$ for its environment \cite{Generalized, MarkovianityDefs, QuantumTrajs}. As we saw previously, the reduced density matrix of S defines how S would be left if an unconditional measurement was performed on its environment. Since in Markovian systems, this can actually be assumed to be happening, the reduced density matrix for S is obtainable by averaging the quantum trajectories for the ensemble of possible bath measurement chains $w(t)$ \cite{MarkovianityDefs,QuantumTrajs}\vspace{-0.11cm}
\begin{equation}
\hat{\rho}_S(t):=tr_{E}[\hat{\rho}_{ES}(t)]=\mathbb{E}_{w(t)}\qty[\ket{\psi_{w(t)}(t)}_S\bra{\psi_{w(t)}(t)}_S]. \vspace{-0.06cm}
\end{equation}
Computationally, this means that if for a given master equation, we obtain the stochastic equation ruling the time evolution of such state-vectors $\ket{\psi_{w(t)}}_S$, we would be able to parallelize the computation of the reduced density matrix by solving several independent "vector equations", instead of a single big "matrix equation" \cite{MarkovianityDefs, QuantumTrajs}. Equations of this kind are the so-called {\em Stochastic Schrödinger Equations} (SSEs) \cite{Generalized, continousMeas}. Note that such a quantum trajectory $\ket{\psi_{w(t)}}_S$ for a Markovian environment can always be physically interpreted in the Copenhagen explanation as a so-called pure unravelling \cite{MarkovianityDefs} (where one would invoke the collapse of the subsystem wavefunction at each $\Delta t$). In the Bohmian view on the other hand, such a quantum trajectory is just a normalized CWF of the subsystem S which is converted into an EWF (thus the normalization), after every significant $\Delta t$.\vspace{-0.05cm}

However, what if we had an environment that got entangled with S, but which never really allowed us to consider an effective collapse? What if the different CWFs of the subsystem S were allowed to interact in any future time, instead of being converted into EWF every $\Delta t$? That is, what if the "quantum trajectories" $\ket{\psi_{w(t)}}_S$ for different $w(t)$ could interact between them in future times, making their time evolutions not independent (and not parallelizable unless approximations are made)? This would mean that "the information leaked" into the environment from S, say, the Bohmian "empty waves", would be able to affect the evolution of the system at any time. Such an environment with "memory" of the entanglement achieved with S could be called a {\em non-Markovian} environment \cite{MarkovianityDefs}. 

It turns out that Bohmian mechanics still allows a "pure state" description for S, since, given the position of the environment ancillas interacting with S,\footnote{To allow non-Markovian SSEs "unravelled" through non-position variables, consider the positions of environment "pointers" coupled with non-position observables of the ancillas around the system. Else, consider the associated unmeasured system information $\B^\psi$ presented in Section 3, or the modal theory corresponding to the unravelled observable.} S has always a CWF, whether the conditioning variables are measured or not \cite{NMisModal, interpretSSE}. In the Copenhagen view, a CWF, does not have a physical interpretation, unless it is an EWF, e.g., unless the conditioning variable is strongly measured. As a consequence, under the Copenhagen view, if a SSE is found for a non-Markovian master equation, the pure state $\ket{\psi_{w(t)}}_S$ at time $t$ can only be understood as the state in which S would be left in if the environment ancillas were strongly measured to give $w(t)$. But, since this would produce a very different subsequent evolution of $\ket{\psi_{w(t)}}_S$, such a measurement can only be seen as a "fiction". Of course, non-Markovian SSEs under the Copenhagen view are still useful as pragmatic computational tools to obtain the reduced density matrix of S. However, dynamical information inherent to each pure state (each CWF), like two-time correlations, should be avoided, unless one accepts some sort of ontological reality (independent of measurement) for the conditioning property of the environment, such as the one provided by the Bohmian theory \cite{NMisModal, interpretSSE}.\vspace{-0.1cm}

This narrative in terms of Bohmian CWFs for non-Markovian open quantum systems is not only theoretically insightful, but also a {\em practical} tool to look for reasonable SSEs as we will exemplify now. In the first section, we arrived at an exact system of equations, Eq. \eqref{SE.GJ}, that described the general time evolution of CWFs in arbitrary settings. In principle, in those equations the CWF of the subsystem S and its environment E are coupled at all times, not only between them, but also with the rest of possible CWFs (signature of the non-Markovianity). However, for specific scenarios, we can make educated guesses for the correlation term $\mathfrak{W}$ of Eq. \eqref{G.Bohm} and the classical potential $U$, to generate a SSE for CWFs of the subsystem S (which {\em need} to be independently evolvable to be a valid SSE). Thus, Eq.\eqref{SE.GJ} is a general framework to look for position SSEs. In fact, this equation system is also a detector of non-Markovian behaviour. As long as the CWFs of the subsystem S, $\psi^\xi(\vec{x},t):=\Psi(\vec{x},\vec{y}^{\:\xi}(t),t)$, are described by a $U$ or $\mathfrak{W}$ that depend explicitly on $\vec{y}^{\:\xi}(t)$, the system will be notably non-Markovian.

As an example application of this method, we developed the BITLLES simulator \cite{tdp,Pois,Thz}. In this simulator, we consider a two-terminal nano-scale electronic device operating at high frequencies (in the order of THz), where both the relevant dynamics of the active region electrons and the current measurement times on the reservoirs are in the sub-picosecond range. We consider the active region to be a non-Markovian open quantum system within the language of Eq. \eqref{SE.GJ} \cite{Thz}. The simulator computes the potential $U$ as a solution of the Poisson equation \cite{Pois}, while $\mathfrak{W}$ is modelled by proper boundary conditions \cite{boundary1, Pois} including the correlations between the active region of the device and the reservoirs. Even electron-phonon and electron-photon decoherence effects can be included \cite{eph, Matteo}.

\begin{figure}[h!]
  \centering
  \hspace*{-0.3cm}
   \includegraphics[width=0.71\linewidth]{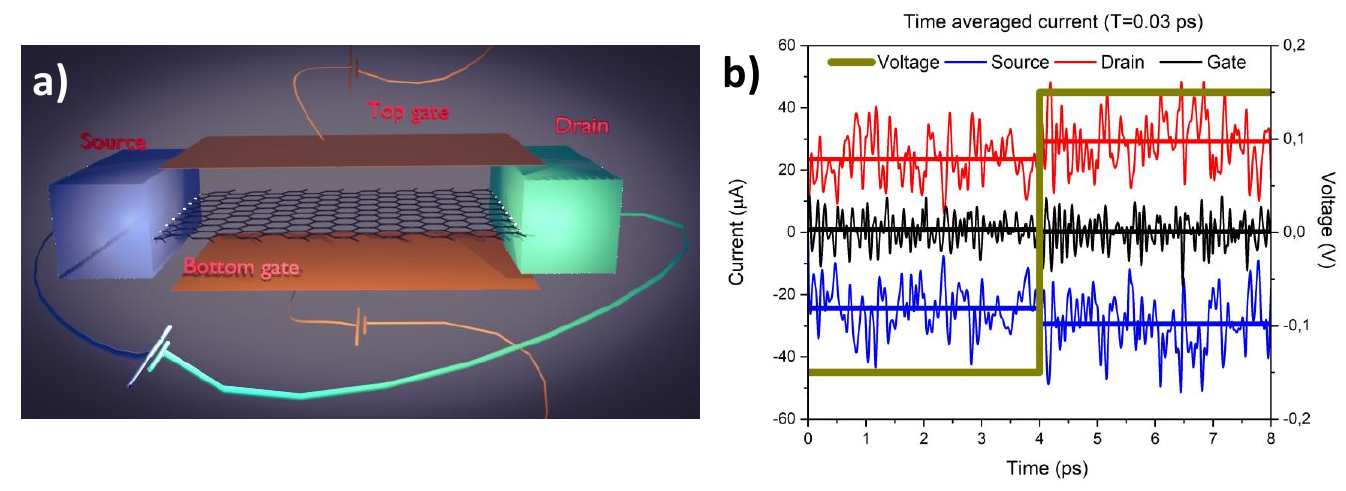}\vspace{-0.3cm}
   \caption{(a) Schematic representation of the graphene-based FET, with a channel composed of a single-crystal mono-layer graphene. (b) The fluctuating lines are the instantaneous currents (time-averaged at a window of $0.03$ ps) as a function of time. The straight lines are due to a wider averaging window of 4 ps, where we can clearly assert the binary response. We can conclude that 4 ps is a reasonable operating time for the transistor.\vspace{-0.2cm}}
  \label{fig:fig}
\end{figure}

In Figure \ref{fig:fig} the ability of the present method is demonstrated \cite{inject,Thz} by predicting for a field-effect transistor (FET) with a graphene channel, the time needed to acknowledge a stable reaction of the drain and source currents when the gate voltage is changed. The Klein tunneling suffered by the electrons while traversing the channel (partition noise) and the random energies of the electrons when injected into the system (thermal noise), cause a fluctuation in the instantaneous current that can be diminished in the laboratory by window averaging. The required window size providing error-free current (averaged) values for digital applications (binary messages) defines the operating frequency of the transistor.\vspace{-0.25cm}

\subsection*{3. Speakable and Operational Information about an Unmeasured system?}
\vspace{-0.08cm}
Returning to the discussion at the beginning of the chapter, under the Copenhagen "eigenstate-eigenvalue link", it is meaningful to say that a quantum system {\em has} a defined property when its wavefunction is an eigenstate of the operator related to that property. Since a strong measurement, as we have seen, always forces the system to adopt an eigenstate, while the unitary evolution in the meantime, will typically cause a superposition, it seems we are only allowed to speak about properties of {\em measured} quantum systems. This makes the predictions about what measurement apparatus pointers show privileged in front of the rest of the information computable using the state of the pre-measurement quantum system. It is true that a quantum theory that correctly predicts what the measurement apparatus pointer will show, is by construction enough for phenomenological predictions. This is why it is argued (even by some Bohmian physicists) that if these predictions are obtainable with empirical agreement, dealing with the rest of the information concealed in the system's state (before its interaction with the measurement apparatus) is just adding unnecessary controversy. However, there are scenarios where the characterization of a quantum system, without the effect of a "collapse backaction" by some measurement pointer, would solve serious {\em practical} difficulties.\vspace{-0.1cm}

As a paradigmatic example, in order to obtain the maximum working frequency of nano-scale transistors (to test the performance of modern computers) \cite{modern}, the time spent by electrons in the active region, their {\em dwell time}, should be provided. The eigenstate-eigenvalue link would force us to place position detectors in the two ends of the active region. However, the quantum measurement, no matter how weak it is, introduces an effective "collapse" backaction in the system that disrupts its future evolution. Thus, the number given by these detectors would be meaningless to benchmark "unmeasured" transistors: no computer has position detectors at the ends of its transistors \cite{tunnel1, tunnel2}. In general, two-time characterization attempts of other "unmeasured" quantum systems face this same problem. For example, in thermodynamics, because {\em work} is by definition a dynamical property implying knowledge of the system (at least) at two different times, it seems there is no possible measurement-context-free definition for a quantum work operator \cite{nogo, workPb1, workPb2}. More generally, {\em two-time correlations} of non-commuting observables, say $F$ and $B$, cannot be defined within the Copenhagen school without including an explicit disturbance by a particular measurement scheme. For example, correlating the result of a strong measurement of $F$ at time $t_1$ and a strong measurement of $B$ at time $t_2$, clearly conveys the disturbing backaction of the measuring device, which collapses the state at $t_1$.\footnote{ An alternative definition could be the real part of the (complex) expectation $\langle \hat{B}(t_2)\hat{F}(t_1)\rangle$ in the Heisenberg formalism, which turns out to be the correlation of a weak measurement \cite{Weak} of $F$ at time $t_1$ and a strong measurement of $B$ at time $t_2$. Yet, as shown in Ref. \cite{DevInPosition2}, even an ideally weak measurement is in fact contextual (in the sense of footnote 12).} Thus, are we fundamentally forbidden to access dynamical information about the "unmeasured system"\footnote{A system that is not being measured, e.g., a closed system evolving without quantum interaction with its environment.}? Or is there a way to consistently define non-{\em contextual}\footnote{Contextual means it depends and implies the particular environment used to convey the information to the observer.} properties (without contradicting the Bell-Kochen-Specker (BKS) theorem \cite{Mermin})? Bohmian mechanics, with its ontology of reality being persistent even when no measurement is taking place, appears to be the escape route. But is it?  \vspace{-0.15cm}

{\em Three impasses} need to be clarified here. First: is there a (Bohmian) way to meaningfully talk about "unmeasured system" features, and even still be in accordance with phenomenology? Second: are these "unmeasured system" features experimentally accessible? If so, how can they be in agreement with the BKS theorem against non-contextual hidden variables? And third: can these features be employed to operationally compute practical information, or are they mere "philosophical reliefs"?
\vspace{-0.25cm}

\subsubsection*{3.1. Breaking Impasse 1: Speakable information of the "unmeasured" system}\vspace{-0.1cm}
Let us first clarify whether the information we obtain by measuring a quantum system is about the pre-measurement/"unmeasured" system or the post-measurement system. Consider an observable $B$ of related operator $\hat{B}=\sum_b b\ket{b}\bra{b}$, with $\{\ket{b}\}_b$ an orthonormal basis and $\ket{\psi}$ the wavefunction of the pre-measurement system. We have seen that (strongly) measuring $B$ will lead the system to the {\bf post}-measurement state $\ket{b}$ linked with the measured (eigen)value $b$, which will happen with a probability $|\bra{b}\ket{\psi}|^2$ due to the {\bf pre}-measurement state. Thus, a single measurement tells us barely nothing about the {\bf pre-}measurement system. But if a "measurement", as Bell pointed out \cite{Bell}, has the connotation of revealing information about the ({\bf pre-}measurement) system, it seems that it would be more proper to name this process an "experiment" rather than a "measurement". We can try to recover the name "measurement" with an ensemble of these "experiments" over identically prepared pre-measurement states $\ket{\psi}$. With them, we could estimate the (squared) magnitudes of the {\bf pre}-measurement projection-coefficients to each eigenstate $|\bra{b}\ket{\psi}|^2$ (e.g., using relative frequencies). Then, one could compute the expectation $\bra{\psi} \hat B \ket{\psi}=\sum_b b |\bra{b}\ket{\psi}|^2$, which is also a number dependent on the pre-measurement state $\ket{\psi}$. However, from a Copenhagen point of view, this number (say, the average energy or position of an electron) can only be interpreted as a property of the {\bf post}-measurement system, because by the eigenstate-eigenvalue link, only the post-measurement system can be attributed the observable $b$. When it comes to Bohmian mechanics, if $\hat{B}$ commutes with position $\hat{x}$, because the position $x$ is "speakable" at all times, the number $\bra{\psi} \hat B \ket{\psi}$ is the average property $B$ of the {\bf pre}-measurement system (as the simplest example, if $\hat{B}=\hat{x}$, it is the average Bohmian position of the unmeasured system). Yet, if $\hat{B}$ does not commute with $\hat{x}$ (e.g., for the momentum or the Hamiltonian operators), it is unclear if the expectation $\bra{\psi} \hat B \ket{\psi}$ computed with the measured $b$ is understandable to be a property of the pre-measurement system. In trying to clarify this, by linking the observable $B$ to the position $x$ of the Bohmian trajectories, which are "speakable", one can find a solution to the first impasse.\vspace{-0.05cm}

Given an arbitrary (Hermitian) operator $\hat{B}$, describing the observable $B$ for the subsystem S, with normalized EWF $\ket{\psi(t)}$, let us define the position function $C^{\psi}(\vec{x},t):={\bra{\vec{x}}\hat{B}\ket{\psi(t)}}/{\bra{\vec{x}}\ket{\psi(t)}}$. If we write the expected value for $\hat{B}$ as a function of $C^\psi(\vec{x},t)$, we get that\vspace{-0.15cm}
\begin{equation}\label{C}
\langle \hat{B}\rangle(t)= \bra{\psi(t)}\hat{B}\ket{\psi(t)}=\int \bra{\psi(t)} \ket{\vec{x}}\bra{\vec{x}}\hat{B}\ket{\psi(t)}dx =  \int |\psi(\vec{x},t)|^2C^\psi(\vec{x},t)dx.\vspace{-0.1cm}
\end{equation}
This means that the spatial average of the (possibly complex) $C^\psi(\vec{x},t)$ yields, at all times, the same expected value for the observable $B$ as that given by the Copenhagen theory. Now, let us define a real function $\B^\psi(\vec{x},t):=\mathbb{R}e\{C^{\psi}(\vec{x},t)\}$. Because $\hat{B}$ is an observable, its expected value will be a real number, such that $\langle \hat{B}\rangle=\mathbb{R}e\{\langle \hat{B}\rangle\}$. Thus, taking the real part of equation \eqref{C}, we get that\vspace{-0.1cm}
\begin{equation}
\langle \hat{B}\rangle(t)=\int |\psi(\vec{x},t)|^2\B^\psi(\vec{x},t)dx.\vspace{-0.1cm}
\end{equation}
We can link this with the set of Bohmian trajectories $\{\vec{x}^{\:\xi}(t)\}_{\xi\in\Sigma}$ sampled in independent repetitions of the experiment, to get that $\langle \hat{B}\rangle(t)=\lim_{|\Sigma|\rightarrow \infty}\frac{1}{|\Sigma|} \sum_{\xi\in\Sigma} \B^\psi(\vec{x}^{\:\xi}(t),t)$,
using the Quantum Equilibrium principle \cite{Absolute}. This means that the real number $\B^\psi(\vec{x}^{\;\xi}(t),t)$, related to the $\vec{\xi}$-th Bohmian trajectory of the "unmeasured" system, when averaged over the ensemble of possible trajectories, gives the same value as the operator's expectation value. That is, irrespective of whether or not we give the observable $B$ an ontological status, we can understand $\B^\psi(\vec{x},t)$ as a mathematical feature related to $B$ and linked to the Bohmian trajectory passing from $\vec{x}$ at time $t$ (in the "unmeasured" system). This is why Holland gave the name {\em local expectation value} to position functions like $\B^\psi$ \cite{Holland}. However, we will just call them the "information linked to $B$ and the Bohmian trajectory at $(\vec{x},t)$" or shortly "information $\B^\psi$", to stress that we (still) mean nothing about their ontological or operational status.

For now, $\B^\psi$ appears to be just an ad-hoc function of the trajectories for the operator expected value to be recovered from those trajectories. Let us see that it is more than this. What would this number be for each trajectory if the system state, $\ket{\psi}$, was an eigenstate of $\hat{B}$ of eigenvalue $b$?
\begin{equation}
\B^\psi(\vec{x})=\mathbb{R}e\qty{ \frac{\bra{\vec{x}}\hat{B}\ket{\psi}}{\bra{\vec{x}}\ket{\psi}} } = \mathbb{R}e\qty{ \frac{\bra{\vec{x}}\ket{\psi}b}{\bra{\vec{x}}\ket{\psi}} }=b\quad \forall \vec{x}.\vspace{-0.05cm}
\end{equation}
This means that a condition for $\ket{\psi}$ to be an eigenstate of $\hat{B}$ is that it is a state for which every Bohmian trajectory has the same value of the information $\B^\psi$. On the one hand, this tells us that the $b$ indicated by the pointer of a projective measurement can always be considered to be information linked to the Bohmian trajectory, even when its operator does not commute with position. On the other hand, in practice, it can be a tool to construct the operator $\hat{B}$ itself. One could define $\hat{B}$ in terms of $\B^\psi$, as the collection of states $\ket{b}$ in which all Bohmian trajectories have the same value $b$ for the information $\B^\psi$ (and some condition to make the imaginary part of $C^\psi$ negligible).

If the explicit shape of $\B^\psi$ had nothing to do with Bohmian mechanics, this reverse definition of $\hat{B}$ would be a pointless definition. However, it turns out that if we set $\hat{B}$ to be the momentum operator $\hat{p_k}$ of the $k$-th degree of freedom, the trajectory information $\B^\psi(\vec{x},t)$ is exactly equal to the Bohmian momentum of the trajectory crossing $\vec{x}$: $m_k v_k(\vec{x},t)$ \cite{DevInPosition1}. If we set $\hat{B}$ to be the Hamiltonian operator $\hat{H}$, the information $\B^\psi(\vec{x},t)$ turns out to be exactly equal to the Bohmian energy (kinetic plus classical and quantum potentials \cite{JordiXavier}) of the trajectory crossing $\vec{x}$. One can see that the list of these "fortunate" matches for position functions that appeared to be designed only to satisfy the expectation values does not end there \cite{Holland}. This suggests that we can employ Bohmian mechanics to derive the expression for $\B^\psi$, thanks to its similarity with classical mechanics, and then define the related operator in those terms. Whether the information $\B^\psi$ has an ontological status or not, whether it is operational or not, this is already (numerically) useful, because there are observables, like the total (particle plus  displacement) current in a nano-device (plotted in Figure \ref{fig:fig}.b), for which there is no clear operator, but there is a clear Bohmian observable associated with it, as will be explained in detail later \cite{Pel, equiv}.

In a nutshell, since we placed no restriction on $\hat{B}$, one is safe to assume that at all times, mathematically, each Bohmian trajectory $\vec{\xi}$, has a simultaneously determined value $\B^\psi(\vec{x}^{\:\xi}(t),t)$ linked to {\em every} observable operator $\hat{B}$. Whether $\B^\psi(\vec{x}^{\:\xi}(t),t)$ reflect an \textbf{ontic property} (a property that the theory postulates to be part of the ontology) or not, that will be given by the quantum theory at hand. For example, we found that when $\B^\psi$ is linked to the momentum operator $\hat{p}_k$, it is equal to the Bohmian particle's momentum, which is an ontic property in Bohmian mechanics, but not in the Copenhagen theory. The key is that when $\B^\psi$ is equal to an ontic property, in any theory where a position trajectory exists in the absence of observation, $B$ becomes "speakable" with a well-defined value at all times. Importantly however, we saw that the information $\B^\psi$ is an equally well-defined number linked to each trajectory, independently of the ontic character of $B$\footnote{The information $\B^\psi$ will evolve continuously as long as the wavefunction evolves unitarily (which in Bohmian mechanics always does, as we saw). Then, if the system evolves from an eigenstate $\ket{b_1}$ to another $\ket{b_2}$ with eigenvalues $b_1\neq b_2$, $\B^\psi$ will take all the intermediate values not necessarily among the eigenvalues of $\hat{B}$. This suggests an interpretation in which the "quantization" of quantum mechanics is only an apparent feature, due to the fact that for a "proper" measurement, we require that a pointer saying $b$ is compatible with a wavefunction $\ket{b}$ that yields for a strong measurement the result $b$ with probability 1. That is, a wavefunction which has all its Bohmian trajectories with value $b$ for $\B^\psi$. Then, we would call it "quantum" because this delicate orchestration can only happen for a certain "quantized number" of wavefunctions (the eigenstates).}. Then, the fact (we will show now) that the $\B^\psi$ can be operationally obtained in a laboratory, will make the information $\B^\psi$ practically useful across the "unspeakables" and independently on the followed theory.
\vspace{-0.2cm}
 
\subsubsection*{3.2. Breaking Impasse 2: Is this "unmeasured" system information operational?}\vspace{-0.15cm}
If we could only obtain the information $\B^\psi$ in a laboratory through a strong von Neumann interaction, hence forcing it to become an eigenvalue of $\hat{B}$, all this would have the same practical limitations as the eigenstate-eigenvalue link (although speaking about these numbers in the absence of measurement would now be "possible"). If so, the information $\B^\psi$ would not be an {\em operational property}\footnote{A number that can be obtained in a laboratory with a well-defined protocol.} of the unmeasured system. However, it turns out that the "unmeasured" $\B^\psi$ is actually experimentally determinable even for a non-eigenstate pre-measurement system. The "how", explains the "cumbersome" definition $\B^\psi(\vec{x},t)=\mathbb{R}e\{\bra{\vec{x}}\hat{B}\ket{\psi}/\bra{\vec{x}}\ket{\psi}\}$. It turns out to be the protocol that naive classical experimentalists \cite{WisemanVel} would follow if they thought the system had a defined position, initially uncertain to them, and the only quantum knowledge they had was that measurement interactions spoil the system's natural subsequent evolution. In order to know the property $B$ of such a subsystem S (say, an electron) when it crosses $\vec{x}$, they would first couple an ancilla A to the subsystem S of EWF $\ket{\psi}$, through the measurement Hamiltonian $\bar{\mu}(t)\,\hat{p}_A\otimes\hat{B}$ but let the interaction strength $\mu$ be very small, such that the system state is only slightly perturbed. They would strongly measure the slightly entangled ancilla's position $z_B$ with a certain probability density $P(z_B)$, getting a weak measurement about the property $B$ of S. Before the slightly perturbed system S further evolved, they would strongly measure its position $z_x$, with a certain conditional probability density $P(z_x|z_B)$. Finally, they would average the weak measurements of $B$ for which the system S (the electron) was found at $\vec{x}$, in order to erase the noise introduced by the weakness of the coupling with A. If the averaged ensemble is large enough, the resulting conditional expectation will be equal to $\int z_B P(z_B|z_x)dz_B$, which as proven in \cite{DevInPosition2}, is roughly equal to $\B^\psi(\vec{x},t)$. This is called a {\em position post-selected weak value} \cite{Weak}.

A naive experimentalist would not be surprised at all by such a "coincidence". One can consider all this was juggling with results of several observations. But, when the information $\B^\psi$ is an ontic property of the theory, one can legitimately say (under that theory), that the average weak measurements of $B$, for experiments in which the system (the electron) was at $\vec{x}$, gave $\B^\psi(\vec{x},t)$, because whenever the trajectory (the electron) was at $\vec{x}$, it had indeed the property $\B^\psi(\vec{x},t)$. Be that as it may, because in principle we can follow this protocol in a lab for most observables $B$, irrespective of their ontic state, $\B^\psi(\vec{x},t)$ is (almost always\footnote{There is a (quite important) exception. Identical particles are always ontologically distinguishable by their trajectories in Bohmian mechanics. In the laboratory however, there are no means to label each individual particle under many-body wavefunctions with exchange symmetry. In consequence, if we follow our weak value protocol to "measure" the information $\B_{(k)}^\psi:=\mathbb{R}e\{\bra{\vec{x}_1,...,\vec{x}_M}\hat{Id}_{(1)}\cdots\hat{Id}_{(k-1)}\hat{B}_{(k)}\hat{Id}_{(k+1)}\cdots\hat{Id}_{(M)}\ket{\psi}/\bra{\vec{x}_1,...,\vec{x}_M}\ket{\psi}\}$ related to the observable $B$ of the $k$-th electron, in a system of $M$ electrons of positions $\vec{x}_k$ and many-body wavefunction $\ket{\psi}$, what we will get instead is the average: $\sum_{k=1}^M \frac{1}{M}\B^\psi_{(k)}(\vec{x}_1,...,\vec{x}_M)$. Hence, the average $\B_{(k)}^\psi$ for a multi-particle Bohmian trajectory can be operational (say, the sum of the current contributions of the active region electrons, as discussed in the next paragraph), but the individual indistinguishable particle $\B^\Psi_{(k)}$ (like the individual electron current contributions) are not, even if they might be ontic properties within Bohmian mechanics.}) an operational property \cite{DevInPosition1, DevInPosition2}.

Let us clarify the non-contextuality of the information $\B^\psi$. Because the Bohmian position and EWF of a subsystem immediately determine $\B^\psi$ for any observable $B$, this apparently violates the BKS theorem \cite{Mermin}, by which there can be no pre-existing variables that {\em non-contextually} determine the measurement outcomes for all observables (not even only for commuting groups of them). This does not preclude the weak values of the above protocol from being non-contextually pre-determined, because they deal with a different notion of "measurement": the hypotheses of the theorem refer to the Copenhagen quantum measurement (of Section 1), while the above weak value protocol "measuring" non-contextual information $\B^\psi$, is an ensemble average of several Copenhagen quantum measurements, each of which is indeed contextual (the Bohmian description of the measurement apparatus is necessary to determine their individual outcomes \cite{operatorsObservables}). Moreover, as we saw (in footnote 13), the value of $\B^\psi$ for a certain subsystem does {\em not} determine a von Neumann measurement outcome, since it is the coupling Hamiltonian (contextual) that forces the pre-measurement $\B^\psi$ (a priori not even "quantized") to evolve to different ("quantized") eigenvalues of the operator $\hat{B}$. And even still, the weak value protocol, does produce a non-contextual $\B^\psi$ value (through many contextual experiments). This is the reason why one might prefer to regard the post-selected averaging as an {\em uncontextualization protocol}. The clarification would be unncessary, however, if history had preserved the original meaning of the word "measurement" as a {\em protocol} that unveils features of a system, existent before the interaction with the external probes. Unfortunately, according to standard quantum mechanics, as stated by Mermin \cite{Mermin}, "the outcome of a measurement is brought into being by the act of measurement itself".
\vspace{-0.2cm}

\subsubsection*{3.3. Breaking Impasse 3: Is this information useful for a non-Bohmian?}
\vspace{-0.15cm}
Regardless of the followed quantum theory and whether one is ready to accept an ontological status for a certain information $\B^\psi$, its relation with expected values and the definition of the observable operator $\hat{B}$ holds mathematically. This has an important practical application that is also useful for a non-Bohmian. The information $\B^\psi$ can be used to numerically estimate the expected value of observables while avoiding the explicit definition of their formal operators. For this, one could express the observable $B$ in the language of Bohmian mechanics to derive the shape of $\B^\psi(x,t)$, and then get the expected value of the operator $\hat B$ (if there is any), by computing the trajectory ensemble average of $\B^\psi$. For example, this is how we predict the expected total electrical current (including the displacement current) crossing the active region of a two-terminal nano-device operating at high frequencies (THz) in the BITLLES simulator \cite{equiv, Pel}. We can define the contribution to the total current through a surface $\sigma$, due to the Bohmian trajectory of a $k$-th electron $\vec{x}_k^{\:\xi}(t)$ of charge $e$, as $I_k^{(\xi)}(t)=\int_\sigma \vec{J}^{\:(\xi)}(\vec{r},t)\cdot d\vec{s}+\int_\sigma \varepsilon(\vec{r},t)\pdv{\vec{E}^{\:(\xi)}(\vec{r},t)}{t}\cdot d\vec{s}$, where $\varepsilon(\vec{r},t)$ is the dielectric permittivity, $\vec{J}^{\:(\xi)}(\vec{r},t)=e\dv{\vec{x}_k^{\:\xi}(t)}{t}\delta(\vec{r}-\vec{x}_k^{\:\xi}(t))$ is the particle current density, and $\vec{E}^{\:(\xi)}(\vec{r},t)$ is the electric field generated by the electron, (as a solution to Gauss' equation). The sum of these contributions, $I^{(\xi)}(t)=\sum_k I^{(\xi)}_k(t)$, will be the total Bohmian current at the surface $\sigma$ for the $\xi$-th experiment in the ensemble $\{\vec{x}^{\:\xi}(t)\}_{\xi\in \Sigma}$. The phenomenological expectation of a total current operator $\hat{I}$ can then be estimated as the ensemble average of these currents, since by the Quantum Equilibrium principle, $lim_{|\Sigma|\rightarrow \infty}\frac{1}{|\Sigma|} \sum_{\xi\in\Sigma} I^{(\xi)}(t)=\langle \hat{I}\rangle(t)$ (if such an operator $\hat{I}$ exists).

In addition, as we saw, the information $\B^\psi$ is (usually) an experimentally obtainable function that, no matter the followed interpretation of quantum mechanics, {\em characterizes} the pre-measurement wavefunction $\ket{\psi}$. This means the $\B^\psi$ can be pragmatically employed to characterize an unmeasured quantum system, just like a tomography or momentum-postselected weak values are useful to obtain the pre-measurement wavefunction \cite{directWF}, no matter the ontological status or speakability of such a wavefunction. In particular, the operational $\B^\psi$ offer a natural solution to the puzzling search of non-contextuality for the metrics involving two different times \cite{DevInPosition1}.\vspace{-0.1cm}

As a first example, they provide a well-defined non-contextual two-time correlation function for general observables. Consider a big set of trajectories $\{\vec{x}^{\:\xi}(t)\}_{\xi\in \Sigma}$ sampled from the pre-measurement wavefunction $\ket{\psi(t)}$. Given the observables $B,F$, the $\xi$-th trajectory has associated informations $\B^\psi(\vec{x}^\xi(t),t):=\R e\{\frac{\bra{\vec{x}^{\:\xi}(t)}\hat{B}\ket{\psi(t)}}{\bra{\vec{x}^{\:\xi}(t)}\ket{\psi(t)}} \}$ and $\mathcal{F}^\psi(\vec{x}^\xi(t),t):=\R e\{\frac{\bra{\vec{x}^{\:\xi}(t)}\hat{F}\ket{\psi(t)}}{\bra{\vec{x}^{\:\xi}(t)}\ket{\psi(t)}} \}$, which are well-defined even if the associated operators $\hat{B},\hat{F}$ do not commute. This gives a natural correlation function defined as \vspace{-0.12cm}
\begin{equation}
\langle B(t_2)F(t_1)\rangle := \lim_{|\Sigma|\rightarrow \infty}\frac{1}{|\Sigma|} \sum_{\xi\in\Sigma} \B^\psi(\vec{x}^{\:\xi}(t_2),t_2)\mathcal{F}^\psi(\vec{x}^{\:\xi}(t_1),t_1).\vspace{-0.12cm}
\end{equation}
 In a similar way, we can solve the problems concerning a quantum work definition, just as done by Refs. \cite{work1, work2}. First note that given a subsystem Hamiltonian $\hat{H}=\sum_k \frac{-\hbar^2}{2m_k}\pdv[2]{}{x_k}+V(\vec{x},t)$, the associated information $\B^\psi$ is \vspace{-0.15cm}
\begin{equation}
\mathcal{H}^\psi(\vec{x}^{\:\xi}(t),t) := \mathbb{R}e\qty[ \frac{\bra{\vec{x}^{\:\xi}(t)}\hat{H}\ket{\psi(t)}}{\bra{\vec{x}^{\:\xi}(t)}\ket{\psi(t)}} ] = \sum_{k=1}^n\frac{1}{2}m_kv_k(\vec{x}^{\:\xi}(t),t)^2+V(\vec{x}^{\:\xi}(t),t)+Q(\vec{x}^{\:\xi}(t),t),\vspace{-0.15cm}
\end{equation}
with $Q$ the well-known Bohmian quantum potential \cite{Holland, Durr, JordiXavier}. This proves $\mathcal{H}^\psi(\vec{x}^{\:\xi}(t),t)$ is, as anticipated, the total Bohmian energy of the $\vec{\xi}$-th trajectory at time $t$. We can compute its associated Bohmian work with $\mathcal{W}^{(\xi)}(t_1,t_2)= \int_{t_1}^{t_2}\dv{\mathcal{H}^\psi(\vec{x}^{\:\xi}(t),t)}{t}dt=\mathcal{H}^\psi(\vec{x}^{\:\xi}(t_2),t_2)-\mathcal{H}^\psi(\vec{x}^{\:\xi}(t_1),t_1)$. As a result, a well-defined non-contextual definition of the quantum work could be the ensemble average of the trajectory works,\vspace{-0.1cm}
\begin{equation}
\langle W(t_1,t_2)\rangle = \lim_{|\Sigma|\rightarrow \infty}\frac{1}{|\Sigma|} \sum_{\xi\in\Sigma} \qty(\mathcal{H}^\psi(\vec{x}^{\:\xi}(t_2),t_2)-\mathcal{H}^\psi(\vec{x}^{\:\xi}(t_1),t_1)).\vspace{-0.1cm}
\end{equation}
Finally, we could give a reasonable Bohmian answer to the pathological search of an "unmeasured" dwell time: the expected time spent by the Bohmian trajectory of the electron within the active region $\Gamma\subset \R^3$. Mathematically, the dwell time $\tau$ for the $\vec{\xi}$-th trajectory of the $k$-th electron with EWF $\psi^\xi(\vec{x}_k,t)$ is by definition given by the integral: $\tau^{( \xi)}= \int_{0}^\infty  dt \int_\Gamma \delta(\vec{r}-\vec{x}_k^{\:\xi}(t)) d\vec{r}$. This makes the expected time $\langle \tau\rangle$ to be given by the Quantum Equilibrium principle as an integral that is already employed to predict the dwell time,\vspace{-0.15cm}
\begin{equation}\label{hauu}
\langle \tau \rangle = \lim_{|\Sigma|\rightarrow \infty}\frac{1}{|\Sigma|} \sum_{\xi\in\Sigma} \tau^{(\xi)} = \int_{0}^\infty dt \int_\Gamma |\psi^\xi(\vec{r},t)|^2d\vec{r}.\vspace{-0.2cm}
\end{equation}
Moreover, it is worth noting that the Bohmian perspective allows to exclude in \eqref{hauu} the contribution of reflected trajecories, providing a more suitable metric for cutoff frequency estimates in electronic devices (since only the transmitted particles have a net contribution to the average electrical current). 

To conclude the section and link it with the discussion on non-Markovian SSEs, notice that because in the non-Markovian case, the trajectory for the "unravelled" environment observable (what we denoted by $w(t)$) can no longer be interpreted as the result of a continuous measurement of the environment, it represents an unmeasured observable of the environment. Thus, this is readily a, perhaps unintended, application of $\B^\psi$-like properties, which happen to be central to simulate the most general quantum systems that interact with many environmental degrees of freedom. \vspace{-0.2cm}

\subsection*{4. Conclusions}\vspace{-0.15cm}
In this chapter, we have seen that inherently Bohmian concepts like the CWF or position post-selected weak values are indeed usable pragmatically as practical tools in the computation of phenomenologically accessible elements, such as the reduced density matrix, expectation values or time correlations. Therefore, with this chapter, we refute the main criticism to the Bohmian theory, by which the trajectories are "unnecessary embellishments" of the orthodox theory, with no practical use. But then, if we can use Bohmian concepts as a tool, why not include them in the standard vocabulary? Not only for their problem-solving utility, but also because they can provide us ontological relief in front of the purely phenomenological Copenhagen view. As we said, this renewed appeal of the Bohmian theory is clearly motivated by a time when no engineer is really capable of accepting the "unspeakable" quantum reality \cite{where, consp}. However, it must be noted that not even great parents of the quantum theory were ready to restrict themselves to the Copenhagen doctrine. For example, regarding the first section, von Neumann in his seminal book \cite{vonNeumann} explains that the collapse law is to be understood as an effective process that should be possible to be considered at an arbitrary point between the subsystem and the macroscopic device, instead of considering it to be a physical phenomenon \cite{NeumannNoCollapse}. Bohr himself assigned the collapse to the contextuality of experimental protocols in terms of macroscopic devices \cite{Dirac}. As we have reviewed, Bohmian mechanics satisfies the claims of both scientists. When it comes to the second section, it was J. M. Gambetta and H. M. Wiseman who pointed out that SSEs for non-Markovian systems tacitly implied the usage of CWFs from modal theories like Bohmian mechanics \cite{interpretSSE, NMisModal} and who suggested the first formal position SSEs for such open quantum systems \cite{WisemanSSE}. Finally, regarding the discussion on the unspeakables of the third section, Dirac himself was an exemplary physicist that employed "unspeakable unmeasured" system properties in the formulation of his major contributions to physics, leaving questioned the "observability doctrine" of the Copenhagen interpretation \cite{Dirac}.

With all this, we might be wondering when will the mainstream decide to break the limiting walls around (non-relativistic) quantum mechanics, as taught to new generations of scientists every day. There is a pedagogical narrative (the Bohmian one) to explain it all while avoiding disjunctives with classical intuitions, a narrative that actually proves to be practically useful by offering additional tools to the Copenhagen theory. Will we someday include it in the standard program of quantum mechanics taught in our universities? Only time will tell.
\vspace{-0.2cm}

{
\printbibliography
}

\end{document}